\documentclass[iop]{emulateapj}
\usepackage{color}
\usepackage{subfigure}

\newcommand{\msun}{M$_\odot$ }
\shorttitle{Do CCCs trigger high-mass star formation? I.}
\shortauthors{Takahira et al.}

\begin{document}

\title{Do cloud-cloud collisions trigger high-mass star formation? I. Small cloud collisions}
\author{Ken Takahira, Elizabeth J. Tasker and Asao Habe}
\affiliation{Department of Physics, Faculty of Science, Hokkaido University, Kita-ku, Sapporo 060-0810, Japan}

\begin{abstract}

We performed sub-parsec ($\sim 0.06$\,pc) scale simulations of two idealised molecular clouds with different masses undergoing a collision. Gas clumps with density greater than $10^{-20}$\,gcm$^{-3}$ ($0.3\times 10^4$\,cm$^{-3}$) were identified as pre-stellar cores and tracked through the simulation. The colliding system showed a partial gas arc morphology with core formation in the oblique shock-front at the collision interface. These characteristics support NANTEN observations of objects suspected to be colliding giant molecular clouds (GMCs). We investigated the effect of turbulence and collision speed on the resulting core population and compared the cumulative mass distribution to cores in observed GMCs. Our results suggest that a faster relative velocity increases the number of cores formed but that cores grow via accretion predominately while in the shock front, leading to a slower shock being more important for core growth. The core masses obey a power law relation with index $\gamma = -1.6$, in good agreement with observations. This suggests that core production through collisions should follow a similar mass distribution as quiescent formation, albeit at a higher mass range. If cores can be supported against collapse during their growth, the estimated ram pressure from gas infall is of the right order to counter the radiation pressure and form a star of 100\,\msun .
\end{abstract}


\section{Introduction}

The role of collisions between the dense clouds of gas in the interstellar medium has flitted on the outside of mainstream star formation theories for over three decades \citep[e.g.][]{Loren1976}. While it is not disputed that such events occur, it has been hard to pin down their frequency and what impact these encounters have on their host galaxy's star formation rate. These problems stem from the difficulty in observing both the collision process itself and its most likely products, the high mass stars. 

The formation process involving the largest members of the stellar population remains one of diverse theoretical candidates. While low mass stars can form via accreting gas onto a gravitationally collapsed core \citep[e.g.][]{Palla1993, Shu1987}, this process most likely fails in the case of a massive star both by the time-scales involved to accrete sufficient mass and by the radiation pressure and ionisation that the growing core would produce, thereby stopping the accretion before a large enough object is formed. To counter these problems, the proposed models either modify the physical conditions of the accretion via such mechanisms as the dust content \citep{Wolfire1987}, wind blown cavities \citep{Krumholz2005} and accretion disks \citep{Yorke2002} or opt for later evolution with physical collisions between low-mass stars \citep{Bonnell1998}. In their paper, \citet{McKee2002} propose that such problems can be avoided if the collapsed core is turbulent and the surface density is sufficiently high with $\Sigma > 0.5$\,gcm$^{-3}$ needed for a $100$\,\msun. In such cases, the ram pressure associated with accretion exceeds the radiation pressure.

The creation of the higher surface densities required by \citet{McKee2002} is a point in favour for cloud-cloud collisions. In this process, interstellar clouds moving at supersonic velocities collide and produce a shock wave at the collision interface. Where the shock compresses the cloud gas, the densities rise and clumps in the post-shocked gas collapse under the effect of the enhanced self-gravity \citep[e.g][]{Stone1970, Gilden1984}. Since clouds moving at the required supersonic velocities are readily observed \citep{Rosolowsky2003, Bolatto2008, Heyer2009} and both the interstellar medium and the star forming Giant Molecular Clouds (GMCs) are known to be turbulent \citep{Larson1981, Elmegreen2002, MacLow2004}, this dynamic picture of triggered core formation has many points in its favour compared to a quasi-static viewpoint. 

\subsection{Observational evidence for cloud-cloud collisions}

Examples of probable cloud-cloud collisions are steadily increasing. In such cases, observations identify two distinct velocity components in the gas, belonging to the two clouds undergoing a collision. Early observations of this phenomenon include \citet{Loren1976, Loren1977} who identified two components in the CO spectra of molecular cloud NGC 133 which they propose is the cause of the chain of newly formed stars within the cloud, and later in LkH$\alpha$\,198, a molecular cloud containing Herbig Be/Ae stars. \citet{Dickel1978} also found two velocity components in Dr21/W75; a high-mass star forming region and \citet{Odenwald1992} accounts the star formation and elongated shape of the unbound high altitude cloud GR110-13 to compression during a cloud collision. 

More recent observations include the two Super Star Clusters, Westerland\,2 and NGC\,3603. Super Star Clusters are tightly packed stellar groups with radii $\sim 1$\,pc and masses $\sim 10^4$\,\msun that contain a large number of massive stars. Their formation has historically perplexed researchers due to the problem of creating such a high mass cluster if the star formation efficiency inside the GMC is low. The first cluster to be identified as a possible collision object was Westerland\,2 which was observed by \citet{Furukawa2009} and \citet{Ohama2010} and found to have two associated clouds with relative velocity of 20\,km/s. The authors propose that the resulting shock between the clouds compressed the gas to produce the massive star cluster. Ohama et al., in prep. later identified two giant molecular clouds associated with NGC3603, prompting suggestions that it could have a similar formation mechanism. A third recent cloud collision candidate was found in observations of the Galactic HII region Trifid Nebula, M20, performed by \citet{Torii2011}. They identified two clouds with masses $10^3$\,\msun and relative velocity $10$\,km/s moving towards a central O star in M20 and suggest this object was created during the clouds' collision. Notably, the relative velocities observed in these three events are too high for the clouds to be gravitationally bound to one another, supporting the idea that these are examples of accidental collisions.

Despite the increasing frequency of observations, the total number of cloud collision candidates remains small, suggesting that such triggered star formation is indeed a minor addition to the Galaxy's total star formation rate. However, new observations using NANTEN2/NANTEN 4\,m sub-mm/mm telescope conversely suggest that such events have previously been miss-classified and are significantly more numerous that previously thought. Torii et al., in prep. argues that the ring structure seen in the Galactic HII region, RCW120, is a classic example of what is expected in a cloud collision. RCW120 is one of the Spitzer bubbles \citep{Churchwell2006}, a collection of 322 objects, 52 of which show a similar ring morphology. This distinctive shape has previously been accounted for by the wind-blown model \citep{Castor1975, Weaver1977}, whereby stellar winds from central high mass star causes gas to be swept into a spherical shell. However, bubbles such as RCW120 display partial arcs, rather than complete rings with high dust concentration on the outer edge of the ring, rather than the interior. The star formation is also off-center, pointing towards a bow shock rather than formation via internal expansion.

\subsection{Theoretical models of cloud-cloud collisions}

The idea that such a morphology is the product of cloud collisions is further supported by theoretical calculations. \citet{Habe1992} performed two dimensional simulations of a head-on collision between non-identical clouds. They found that the larger cloud was disrupted by the resulting bow-shock which in turn, compressed the smaller cloud. This compression caused the post-shock gas in the smaller cloud to become gravitationally unstable, even in the case where the cloud was initially below its Jeans mass. The geometric structure of the collision shows the same ring-like morphology as seen in the Spitzer bubbles, with dense cores for the expected star formation forming in the compressed shock, not in the center of the ring. 

\citet{Anathpindika2010} performed a similar simulation where they looked at the stability of the post-collision bow-shock in three dimensional simulations. While initially supported by turbulence, shocks within the bow-shock cause it to be susceptible to a number of hydrodynamic instabilities including the Kelvin-Helmholtz instability between shearing fluids, the Rayleigh-Taylor buoyancy instability and the Thin Shell Instability that occurs when a shocked shell of gas becomes unstable to small perturbations, developing expanding ripples. Of these three, the Thin Shell Instability was deemed the most important for the bow-shock evolution, causing it to bend at the apex and extend along the collision axis in a filament that fragments to form cores. The creation of such a structure was proposed to be the origin for filament-type clouds, such as the Orion molecular cloud. 

\citet{Anathpindika2010} also note the importance of the relative collision speed between the clouds, finding that a relative velocity of $20$\,km/s produced cores with masses between $25-30$\,\msun in clouds of mass $5000$\,\msun and $2000$\,\msun whereas a lower velocity of $\sim 5$\,km/s failed to produce any cores. This find was further supported by colliding flow simulations of magnetised non-homogeneous gas by \citet{Inoue2013}. They found the mass of the generated core depended both on this collision velocity and the strength of the initial magnetic field.

On a larger scale, \citet{Tasker2009, Tasker2011} have tackled the problem of how common accidental collisions between clouds are likely to be. They find that in a simulation of a Milky Way-type disk, multiple collisions can occur per orbital period, with a rate of approximately 5 per orbit. This agrees with analytical calculations performed by \cite{Tan2000} who argues that --if the rate of collisions is enough to achieve multiple encounters per orbital period-- then cloud collisions could drive the star formation rate in the galaxy, producing the empirical Kennicutt-Schmidt relation between gas surface density and star formation rate \citep{Kennicutt1998}. 

In this paper, we examine the formation of dense, star-forming cores in a head-on cloud cloud collision between non-identical clouds and investigate the impact of turbulence and collision speed on the properties and evolution of these cores.  Section \S 2 describes the details of our simulation, section \S 3 looks at the effect of turbulence, section \S 4 explored the collision speed and in section \S 5 we present our conclusions. 

\begin{figure*}[!t]
\begin{center}
\subfigure{\includegraphics[width=1.0\textwidth]{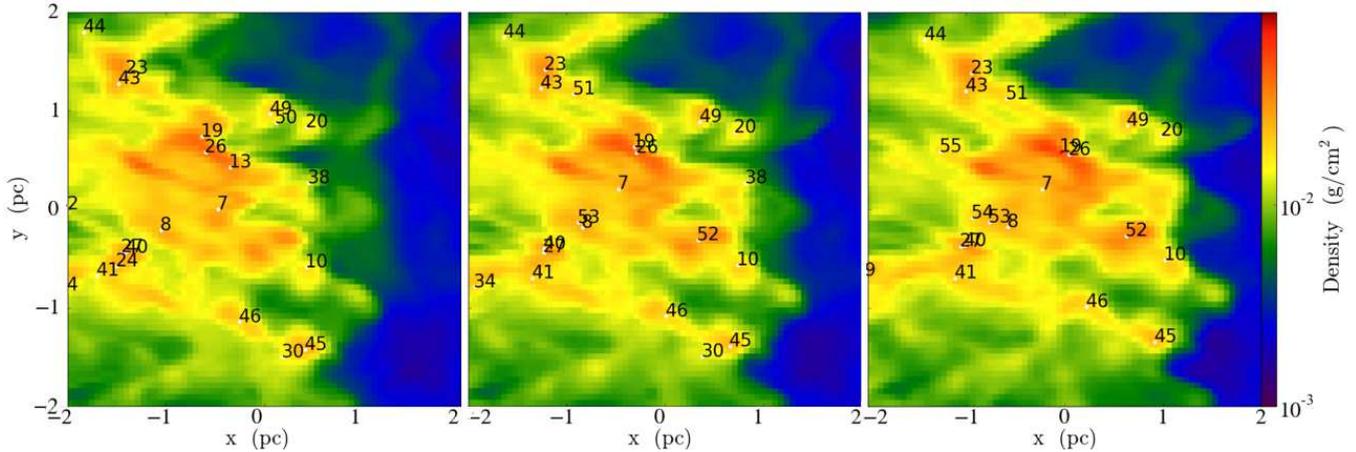}}
\caption{Time evolution of the surface density in a 4\,pc patch of the shock front during a collision. The relative velocity of the colliding clouds shown is 5\,km/s and the time difference between each frame is 0.1\,Myr. White circles mark the positions of the identified cores and the number is the associated tracking value that stays with the core during its lifetime. }
\label{fig:trace}
\end{center}
\end{figure*}

\section{Numerical methods}

Our simulations were run with {\it Enzo}; a three-dimensional hydrodynamical adaptive mesh refinement code \citep{Enzo2013, Bryan1999, Bryan1997}. We used a boxsize of side $30$\,pc with a top level root grid size of $128\times 128\times 128$ and an additional two levels of refinement, giving a limiting resolution (smallest cell size) of $\sim 0.06$\,pc. The refinement criteria was based on baryon mass and on the resolution of the Jeans Length, which must be refined by at least four cells as suggested by \cite{Truelove1997}. 

The hydrodynamics were evolved over time using a three-dimensional implementation of the {\it Zeus} astrophysical code \citep{Stone1992}. To handle shocks, {\it Zeus} uses an artificial viscosity where the associated variable, the quadratic artificial viscosity term, set to $2.0$ (the default value). 

Gas cools radiatively down to $10$\,K using a one-dimensional cooling table created using the CLOUDY cooling code \citep{Ferland1998} where solar metallicity and a density of 100\,cm$^{-3}$ was assumed. For the densities achieved in our simulation, the cooling function remains relatively constant, allowing us to use this simplification. 

In order to prevent gas undergoing unrefined collapse on the finest grid level, we impose a pressure floor. In a collapsing region where the Jeans length can no longer be refined by the desired four cells, the pressure is increased so that the gas follows a polytrope with an adiabatic index $\gamma = 2.0$. This polytrope is stable and stops the collapse at a finite density, preventing single cells getting impossibly dense compared to their surroundings as gravitationally unstable gas flows towards their center. 

While the Truelove limit of four cells per Jeans length is designed to prevent spurious fragmentation, it may not be sufficient to resolve the turbulence on the smallest scales. In simulations performed by \citet{Federrath2011}, 30 cells per Jeans length were required to converge the values of the turbulent energy. In our current simulations, this cell number corresponds to a physical length greater than our core size. We therefore note that while the overall fragmentation of the gas into star-forming regions is correctly followed, the velocity dispersion within these cores is unlikely to have reached a converged value. However, since the primary questions in this paper focus on the formation of potential star forming sites, this limit is sufficient to assess the impact of the cloud collision. To confirm this, in section~\,\ref{sec:highres} we compare our results with a simulation performed at twice the resolution. 

A related point with the above is the impact of the pressure floor on the dense cores. The increased pressure at our maximum refinement prevents unresolved collapse, but also impacts the evolution of the internal core properties. An alternative technique would have been to utilise sink particles which estimate an accretion rate according to the surrounding density and temperature when the gas reaches a high density and meets additional criteria that identify bound and collapsing regions \citep[e.g.][]{Federrath2010}, however due to where our resolution becomes limited and the difficulty with destroying a sink, we chose to keep the cores as pressure supported gas, but note this introduces an error in their internal values.

\subsection{Core Identification and tracking}
\label{sec:setup_cores}

Star-forming cores are identified in the simulation via a constant density contour finding algorithm \citep{yt}. In the majority of our results, this density threshold is set to $\rho_{\rm crit} = 10^{-20}\,{\rm g cm^{-3}}$, although we also compare the evolution in core number with core selection at $\rho_{\rm crit} = 10^{-19}\,{\rm g cm^{-3}}$ and $\rho_{\rm crit} = 5 \times10^{-21}\,{\rm g cm^{-3}}$ in section\,\ref{sec:core_number}. In comparison, the average GMC density is observed to be $\sim 100$\,${\rm cm^{-3}}$ while star formation is seen to occur at $10^4$\,${\rm cm^{-3}} \simeq 3\times 10^{-20}\,{\rm g cm^{-3}}$ (observations by \citet{Ginsburg2012, Lada2010} and theoretical estimate from \citet{Padoan2013}). Therefore within these cores, gravity is likely to dominate over other physical processes to result in star formation.

In order to analyse the evolution of the core properties, we track their motion over the time of the simulation. This process is performed in a similar manner to the cloud tracking scheme presented in \citet{Tasker2009} and works as follows:

\begin{enumerate}
\item Data outputs are taken at 0.1\,Myr intervals during the simulation and cores are identified via the above criteria in all outputs.
\item The position of each core is predicted in the next time step via $r_{\rm c, t} = r_{\rm c, t-1} + v_{c, t-1}\Delta t$, where $r_{\rm c, t}$ is the core position at time $t$, $r_{\rm c, t-1}$ and  $v_{c, t-1}$ are the core position and velocity one output earlier and $\Delta t$ is the time passed between outputs, 0.1\,Myr. 
\item A spherical volume of radius $2v_{c, t-1}\Delta t$ is searched around each predicted position for cores present in the time, t output. The closest core in that region is assumed to be the same core 0.1\,Myr later. If no cores are found, the core is assumed to have dissipated or merged with another core. 
\item In the case of two cores being assigned to the same present-time core, the closest one is tracked.
\end{enumerate}

An example of the core tracking is shown for three times at 0.1\,Myr intervals  in Figure~\ref{fig:trace}. The surface density of a 4\,pc section of the shock front during a collision with a relative velocity of 5\,km/s is imaged with the core positions and their assigned number overlaid. By eye it is possible to confirm that cores found in equivalent positions as the shock progresses maintain the same tracking number. 

\section{The Initial Conditions}
\label{sec:ics}

\begin{figure}
\centering
\includegraphics[width=6cm]{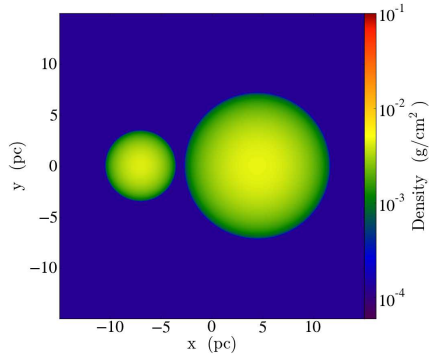}
\caption{Surface density of the initial conditions. Both spheres have a Bonnor-Ebert density profile and constant temperature. When turbulence is included, the velocity field is perturbed as described in section~\ref{sec:ics}.}
\label{fig:ics}
\end{figure}

The simulation initial conditions consist of two clouds of unequal masses whose boundaries are separated by approximately 1.2\,pc  as shown in Figure~\ref{fig:ics}. The density distribution of each cloud takes the form of a Bonner-Ebert sphere \citep{Bonnor1956}; a hydrostatic isothermal self-gravitating gas sphere that is confined by its external pressure. The maximum mass such a sphere can have and remain stable is given by:

\begin{equation}
M_{BE} = \frac{c_{\rm BE} c_s^4}{P^{1/2}_{\rm ext}G^{3/2}},
\end{equation}

\noindent where $c_s$ is the isothermal sound speed, $c_{\rm BE} = 1.18$ is a constant, $P_{\rm ext} = 4000 \times k_B$ ($k_B$ : Boltzmann constant) is the external gas pressure and G is the gravitational constant. Both the simulated clouds fulfil this stability requirement, having masses of $417$\,\msun and $1635$\,\msun, both equivalent to $0.43$\,M$_{\rm BE}$ at 120\,K and 240\,K respectively. Their properties are detailed more fully in Table\,\ref{table:ics}. Once cooling begins, the cloud becomes unstable and without additional support, will start to collapse. 

This profile was primarily chosen because the initial stability is appealing when considering the impact of outside effects. However, molecular clouds have also been observed to show a Bonner-Ebert density profile in the work of \citet{Alves2001}. Other clouds, meanwhile, have a higher mass than their Bonner-Ebert mass but despite this are not in free-fall collapse, due to additional support from internal turbulent motions. This allows $c_s$ in the above equation to be switched for the effective sound speed that also including a contribution from the velocity dispersion. 

Compared to observed GMCs, the cloud mass we have selected is small. Typical clouds within the Milky Way have masses $\sim 10^4-10^5\,{\rm M}_\odot$ \citep{Heyer2009}, a factor of 10 above our chosen sizes. The results from collisions between larger clouds will be explored in subsequent papers, but selecting a smaller cloud at this stage allows a simpler study of the shock front evolution at high resolutions. 

\begin{table}[htdp]
\scalebox{1.5}[1.5]{
\begin{tabular}{ l| l l } 
& Cloud 1 & Cloud 2 \\[0.5ex] \hline
$T_{\rm BE}$\,[K] & 120 & 240  \\[0.5ex] 
$t_{\rm ff}$\,[Myr] & 5.31 & 7.29 \\[0.5ex] 
$r_c$\,[pc] & 3.5 & 7.2 \\[0.5ex] 
$M_c$\,[\msun] & $417$ & $1635$ \\[0.5ex] 
$\sigma_v$\, [km/s] & 1.25 & 1.71 \\[0.5ex] 
$\bar n$\,[${\rm cm^{-3}}$] & 47.4 & 25.3\\[0.5ex] 
k-mode & 6-12 & 10-25 \\[0.5ex] 
\end{tabular}
}
\caption{Initial cloud parameters. From top to bottom: temperature, free-fall time, radius, mass, velocity dispersion, average density and initial k-modes used when turbulence is included.}
\label{table:ics}
\end{table}

\subsection{Turbulence}
\label{sec:turbulence}

In simulations where the gas is given an initial turbulence, we impose a velocity field with power spectrum $v_k^2 \propto k^{-4}$, corresponding to the expected spectrum given by Larson for giant molecular clouds \citep{MacLow1998, Larson1981}. To ensure the turbulence modes were adequately resolved, we selected a maximum $k$-mode value of $1/10$th of the number of cells across the cloud. In addition, we removed the lower order modes since these larger-scale perturbations disrupted the cloud structure, causing it to fragment prior to collision. In previous work designed to model turbulence in giant molecular clouds, driving on the larger scales was found to produce a closer match to observations \citep{Heyer2004, Brunt2004, Brunt2009}. However, in our case we want the focus of the results to be on the impact of the cloud collision, so our choice of turbulent modes is dominated by those that would stabilise the cloud preventing collapse prior to collisional contact from the gas cooling. For our smaller cloud, {\it Cloud 1}, we selected $6 < k < 12$ while the larger cloud, {\it Cloud 2}, was given $10 < k < 25$. The amplitude of the turbulence was dictated by the Mach number, $\cal{M}$ $\equiv \sigma/c_s$, where $\sigma$ is the velocity dispersion inside the cloud and $c_s$ is the sound speed. Prior to the cloud being given a bulk velocity, the $\cal{M}$ $= 1$. The effect of the turbulence on the clouds' stability is discussed further in section~\ref{sec:static}.

When turbulence is applied, the clouds remain in their static positions for $0.5$\,Myrs. This allows the clouds to reach a new equilibrium stage with the turbulent support, as measured by their volume density distribution evolving to the expected lognormal profile for super-sonic isothermal turbulent gas \citep{Vazquez1994, Scalo1998, Ostriker1999} (note, the cloud can be considered isothermal prior to collision since this time for static evolution is approximately 50 times as long as the cloud's cooling time, causing the clouds to reach 10\,K within 0.1\,Myr). After this time, {\it Cloud 1} is given a bulk velocity in the direction of {\it Cloud 2}. Where no turbulence is included, the motion begins immediately. 

\subsection{Static Evolution of Clouds}
\label{sec:static}

\begin{figure}[!t]
\begin{center}
\includegraphics[width=1.0\columnwidth]{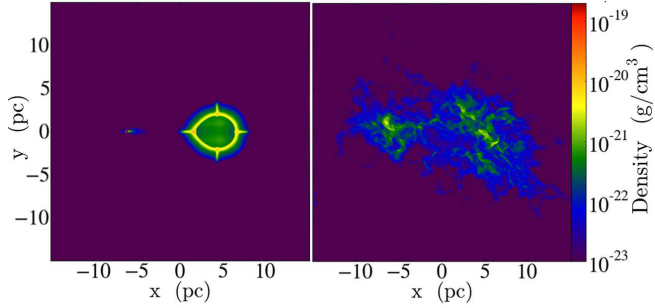}
\caption{Density slice of the two clouds after one complete free-fall time of {\it Cloud 1}, 5.31\,Myr, for static (non-colliding) simulations. Left image shows the clouds where no initial turbulence has been applied while the right image shows the outcome of adding an initial turbulent spectrum to both clouds. Without turbulence, the clouds collapse with {\it Cloud 1} forming a dense point at $x =-7$\,pc. When turbulence is included, the two clouds remain supported.}
\label{fig:image-noturb-turb}
\end{center}
\end{figure}

\begin{figure}[!t]
\begin{center}
\includegraphics[width=1.0\columnwidth]{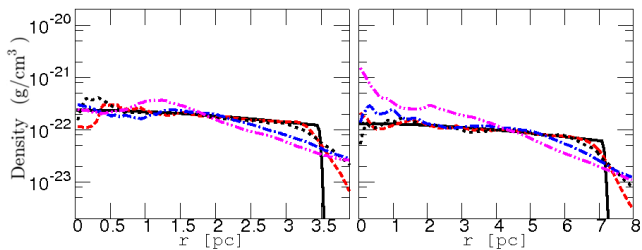}
\caption{Evolution of {\it Cloud 1} (left) and {\it Cloud 2}'s density distribution in the absence of any collision when turbulence is included. The black solid line shows the initial distribution, the red dashed line marks the distribution at 0.5\,Myr, the green dotted line is the distribution at 1.0\,Myr, the blue dot-dash line is at 3.0\,Myr and the pink dot-dot-dash line at 5.3\,Myr, one free-fall time for {\it Cloud 1}.}
\label{fig:profile-static-turb}
\end{center}
\end{figure}

\begin{table*}[!thdp]
\begin{center}
\begin{tabular}{c|ccccccc} \hline
& Turbulence & $v_{\rm coll}$ & $\Delta$ x & Shock speed & $t_{\rm cross_1}$ & $t_{\rm cross_2}$ & Mach number \\ \hline
noTurb & no & 3.0km/s & 0.058pc &2.0 km/s & 5.5Myr & 6.54Myr & 12.25 \\[0.5ex] 
3CC & yes & 3.0km/s & 0.058pc &2.1 km/s & 5.7Myr & 7.82Myr  & 11.22 \\[0.5ex] 
5CC & yes & 5.0km/s & 0.058pc& 2.9 km/s & 4.5Myr & 3.22Myr & 18.7 \\[0.5ex] 
10CC & yes & 10.0km/s & 0.058pc & 7.5 km/s & 1.8Myr & 2.76Myr & 37.4 \\[0.5ex] 
5CCHR & yes & 5.0km/s & 0.029pc & 2.4 km/s & 4.3Myr & 2.67Myr & 19.46 \\ \hline
\end{tabular}
\caption{Summary of the simulations performed. $v_{\rm coll}$ is the relative speed of the two clouds, $\Delta x$ is limiting resolution in each simulation and $t_{\rm cross_1}$ and $t_{\rm cross_2}$ are the shock crossing times across {\it Cloud 1} and {\it Cloud 2}. The Mach number is defined by $v_{\rm coll} / c_s$, where $c_s$ is the sound speed.\label{table:runs}}
\end{center}
\end{table*}

Before considering the cloud-cloud collision, the duration of the clouds' stability is tested while in-situ over the free-fall time of {\it Cloud 1} (see Table~\ref{table:ics}). This time is longer than the collision duration (as measured by the shock crossing time) for the two faster collision speeds we consider in section~\ref{sec:results_velocity} and approximately $70$\% of the collision duration for our slowest collision speed.  

This run was performed to differentiate between structural evolution from the collision process and independent changes within the cloud. Figure~\ref{fig:image-noturb-turb} images the density for the case when turbulence is not included (left) and where it is added to the initial conditions as described in Section~\ref{sec:turbulence}. Without turbulence, the clouds collapse as they cool within the expected free-fall time. After 5.3\,Myr, one free-fall time for {\it Cloud 1} has passed and its radius has contracted to the dense point shown at $x = -7$\,pc in Figure~\ref{fig:image-noturb-turb}. When turbulence is included, the cloud is supported against the cooling over the same time duration. 

In the left-hand pane of Figure~\ref{fig:image-noturb-turb}, four dense lines are seen like compass points on {\it Cloud 2} as it collapses. These have the hallmarks of a numerical artefact and are due to a preferential grid alignment in the fluid instabilities. As the gas cools, the cloud's internal energy decreases and pressure equilibrium is broken. As the innermost region of the cloud begins to collapses, the background gas presses down on the surface layer, creating a dense outer ring which becomes unstable to the Rayleigh Taylor and thin shell instabilities \citep{Vishniac1983, McLeod2013}. These initial ripples first develop along the grid direction (as can sometimes happen in Eulerian codes, see for example \citet{Tasker2008}) and get emphasised as the cloud contracts. With the addition of turbulence, this effect becomes negligible and we therefore ignore it in future runs. The role of the thin shell instability is discussed further in section~\ref{sec:results}.

The effectiveness of the turbulence is seen more clearly in Figure~\ref{fig:profile-static-turb}, which plots the radially averaged density profile for {\it Cloud 1} (left) and {\it Cloud 2} over 5.3\,Myr when turbulence is included. The black line shows the initial profile for each cloud while red-dashed, green-dots, blue-dash-dot and pink-dot-dot-dash show the profile at 0.5\,Myr, 1.0\,Myr, 3.0\,Myr and 5.3\,Myr. In the actual collision simulations, 0.5\,Myr marks the start of {\it Cloud 1}'s movement towards {\it Cloud 2}, 1.0\,Myr is the start of the collision (cloud edges touching) for our slowest, $3.0$\,km/s, collision speed and 5.3\,Myr is the free-fall time for {\it Cloud 1}. Both profiles show only a small degree of evolution, indicating that the turbulence supports the cloud effectively as it cools over this time scale. While there is some formation of structure from the turbulence alone, it is negligible compared to our results during the cloud-cloud collisions in the next section. A core analysis for both clouds in this run finds no cores with density greater than our primary threshold of $\rho_{\rm crit} = 10^{-20}\,{\rm g cm^{-3}}$. While we expect the turbulence to decay over time, this analysis shows that its presence is enough to support the cloud and prevent star-forming cores forming from cloud instability alone over the period of analysis in this paper.

\section{Results}
\label{sec:results}

The collision between {\it Cloud 1} and {\it Cloud 2} in Table~\ref{table:ics} was performed under a variety of different conditions. In the first presented simulation (noTurb), the clouds were given no internal turbulence and collide with a relative velocity of 3\,km/s. The next three simulations begin with the clouds supported against gravitational collapse by turbulence, as described in Section~\ref{sec:turbulence}, and collide with velocities 3\,km/s (3CC), 5\,km/s (5CC) and 10\,km/s (10CC). The final simulation repeats the 5\,km/s event at higher resolution (5CCHR). A summary of these runs is given in Table~\ref{table:runs}. In the table's listed values, $t_{\rm cross_1}$ and $t_{\rm cross_2}$ are the times for the shock to pass through {\it Cloud 1} and {\it Cloud 2} respectively after contact. Both these and the shock speed are measured by plotting the shock progression in the density along a line-of-sight directed along the collision axis. 

\subsection{Non-turbulent collision} 
\label{sec:results_noturb}

\begin{figure}[!th]
\begin{center}
\subfigure{\includegraphics[width=\columnwidth]{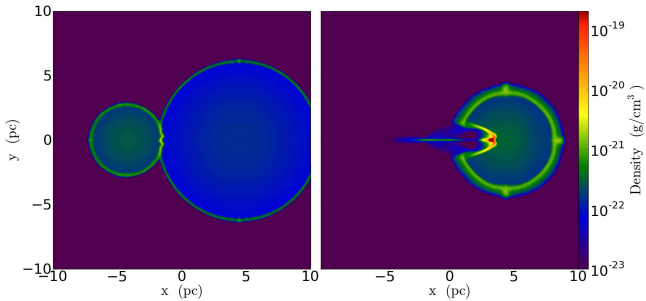}}
\end{center}
\caption{Mid-plane gas density of the cloud collision case without initial turbulence. The left-hand panel shows the two clouds at $t= 1.4$ \,Myr as their boundaries begin to touch. The right-hand panel shows the formation of the single bound core at $t=3.9$ \,Myr. This formation time is shorter than the {\it Cloud 1}'s free-fall time at $t_{ff_1}=5.3$\,Myr, demonstrating that the gravitational instability has been enhanced by the collision process.}
\label{fig:noturb}
\end{figure}

\begin{figure*}[!th]
\begin{center}
\includegraphics[width=1.0\textwidth]{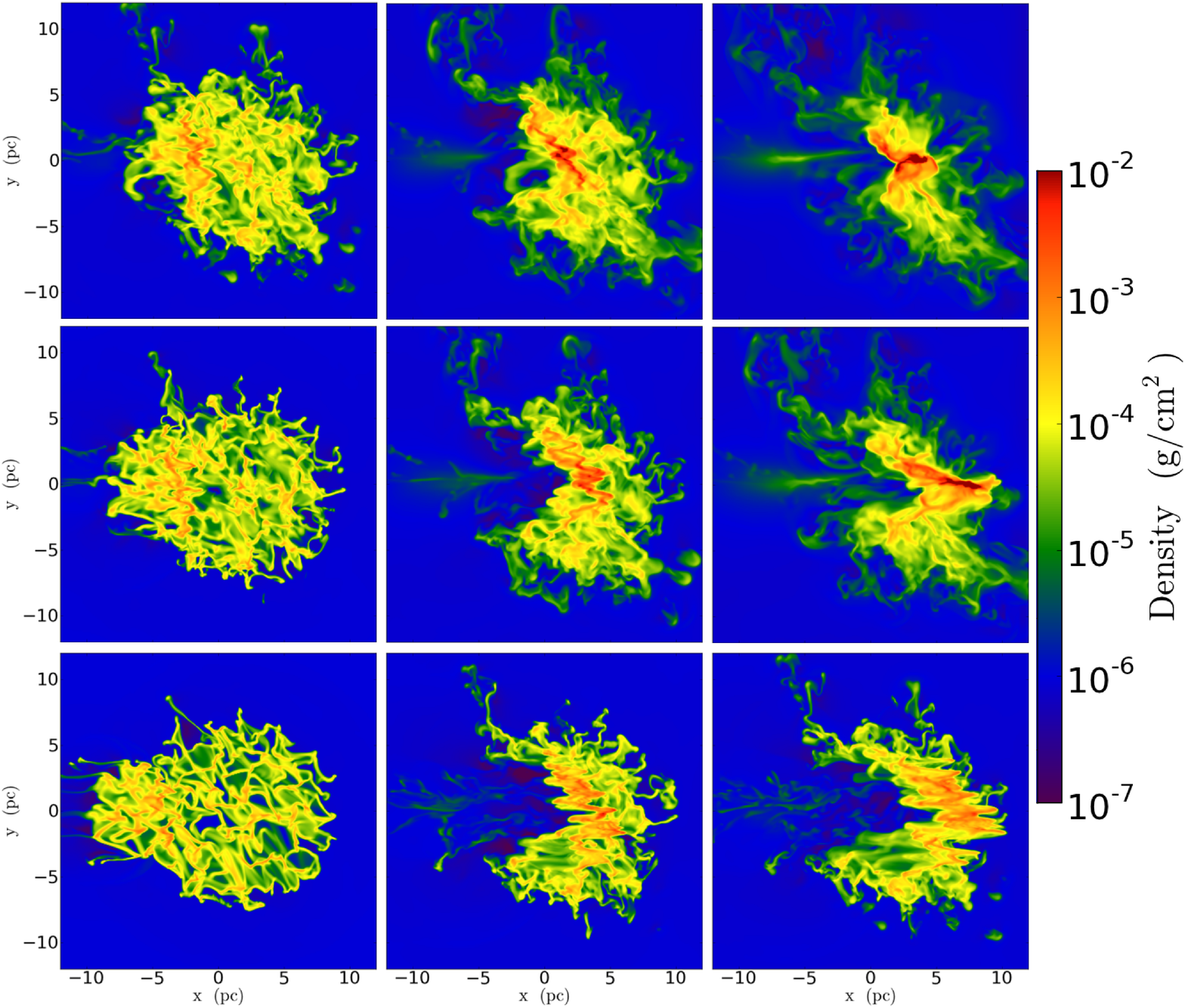}
\caption{Thin (0.2\,pc thickness) projections of the cloud collision at different relative velocities. The three times shown correspond to the first core formation (left), maximum number of formed cores (middle) and when the shock wave has exited {\it Cloud 2} (end of simulation). The top panel shows the simulation with 3.0\,km/s collisional velocity, middle panel shows the results for 5.0\,km/s and the bottom panel is for 10\,km/s.}
\label{fig:collvel}
\end{center}
\end{figure*}

Figure~\ref{fig:noturb} shows density slices from the cloud collision without initial internal turbulence at $t = 1.4$\,Myr (left) and $t = 3.9$\,Myr. The relative velocity of the clouds is 3.0\,km/s. In the left-hand panel, the clouds are shown at the early stage of the collision with their boundaries having just begun to touch. By this point, the clouds have cooled to a gas temperature of 10\,K (the lower limit of our cooling curve), reducing the internal gas pressure and causing the clouds to begin gravitational collapse from their own self-gravity. The outer rims of the clouds are compressed by the external gas pressure, forming thin compressed gas shells at their surface. The shells themselves show a small, irregular ripple structure due to the Rayleigh-Taylor buoyancy instability triggered by their lower density surroundings. 

Strong shock waves appear at the interface between the clouds, forming a thin dense disk of shocked gas. Within this disk, the ripple structure in the thin surface shells is enhanced due to the thin shell instability, with the thin shell being bounded on both sides by the progressing shocks from the collision and accelerated by {\it Cloud 1}'s continual motion. 

Since it has the higher density, {\it Cloud 1} starts to penetrate the larger {\it Cloud 2}, producing an oblique shock front that presses on the concave region of the thin disk, increasing the gas surface density and continuing to push down. Meanwhile, the shockwave entering {\it Cloud 2} becomes a bent bow shock. This is seen clearly in the right-hand panel of Figure~\ref{fig:noturb}, which shows the same collision at the later stage of $t = 3.9$\,Myr. By this time, the shocked thin disc has laterally contracted by its own self-gravity to form a bound dense core at its center. This result agrees well with the previous two-dimensional study performed by \citet{Habe1992} who also found that due to the size difference between the two clouds, the shock wave entering the larger cloud forms bow shock. At its apex, the smaller cloud's post-shocked gas is compressed into a disk which collapses under self-gravity to form a single dense core. This core continues to move through the larger cloud, which simultaneously continues to contract under gravity.

The formation of the bound core occurs at 3.9\,Myr, considerably shorter than {\it Cloud 1}'s free-fall time of $t_{ff_1} = 5.31$\,Myr. The collision process therefore enhances the gravitational instability of the cloud. Since the entire cloud is compressed into this core, its mass is equal to that of {\it Cloud 1}. 

\subsection{The effect of varying collision velocity}
\label{sec:results_velocity}

We move on to exploring the effect of the clouds' relative collision speed on the properties of the cores formed during the interaction. The three simulations presented in this section all use turbulence to initialize {\it Cloud 1} and {\it Cloud 2} which collide with relative velocities 3\,km/s, 5\,km/s and 10\,km/s. As described in section~\ref{sec:turbulence}, the clouds remain stationary for 0.5\,Myr to allow the development of their internal turbulent structure, after which {\it Cloud 1} begins to move towards the stationary {\it Cloud 2}. 

The evolution of these three simulations is shown visually in Figure~\ref{fig:collvel}, with time progressing from left to right. Each image shows the surface density averaged over a depth of $0.2$\,pc. The top row of three panes shows the evolution of the simulation with a relative cloud collision speed of 3.0\,km/s, the middle row shows the results using a collision speed of 5.0\,km/s while the bottom row shows the final tested speed of 10\,km/s. The times shown correspond to the same event in each simulation: the panes furthest on the left show the time of the first core formation (3.0, 1.6 and 0.9\,Myr respectively for the 3\,km/s, 5\,km/s and 10\,km/s simulation after the runs begin), the middle pane shows the time where there exists the maximum number of cores (4.6, 3.7 and 2.2\, Myr) and the final right-hand pane shows the clouds as the shock exits {\it Cloud 2} (6.5, 5.2 and 2.8\,Myr). 

\begin{figure}[!th]
\epsscale{1.2}
\plotone{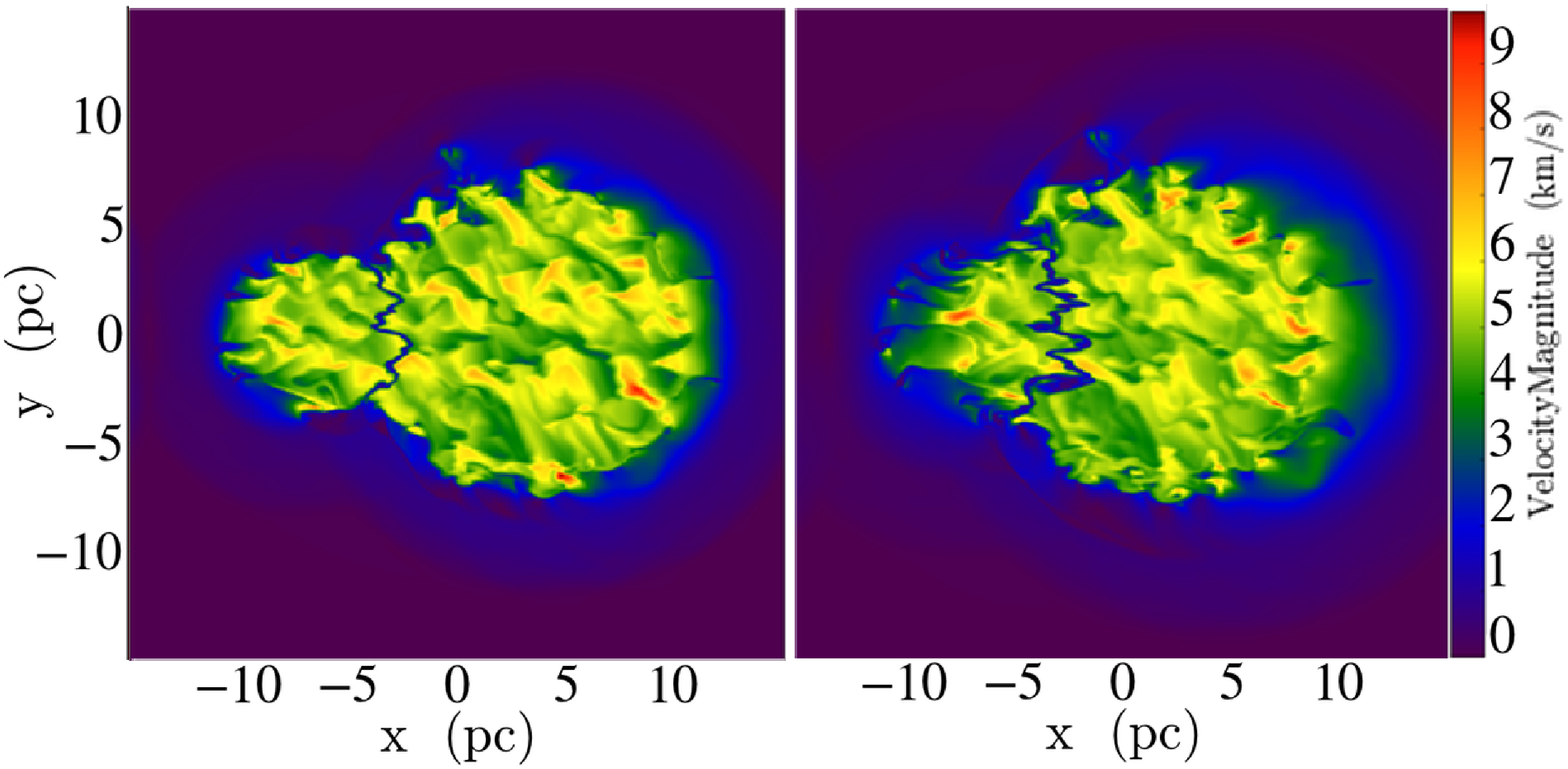}
\caption{Velocity magnitude distribution in the collision between {\it Cloud 1} and {\it Cloud 2}, with both clouds moving prior to collision with velocities 5\,km/s and -5\,km/s. This is equivalent to our 10.0\,km/s case, but because both clouds move, the form of the shock front can be seen clearly. This simulation gives identical results to the 10\,km/s simulation and is included here only to aid the description of the shock front formation. The left-hand pane shows the clouds at $t = 0.8$\,Myr while the right-hand pane shows the slightly later time, $t = 1.2$\,Myr.}
\label{fig:shockfront}
\end{figure}

In all three cases, the initial progression of the shock formed in the collision is the same: Since the relative velocities of the colliding clouds are supersonic, isothermal shock waves appear in both clouds as was seen in section~\ref{sec:results_noturb} for the non-turbulent collision case. However, since the pre-shocked gas in the clouds now has a filamentary structure formed from via the turbulence, the propagation of the shock becomes more complicated. 

In Figure~\ref{fig:shockfront} we show an alternative simulation to the main ones presented in this section. The relative velocity between the two colliding clouds is 10\,km/s, but unlike our previous 10\,km/s case, both the clouds move with velocities 5\,km/s and -5\,km/s. The results from this are identical to that with {\it Cloud 1} moving with 10\,km/s, but the shape of the shock front is now starkly visible in the magnitude of the gas velocity that is shown in Figure~\ref{fig:shockfront}. In the left-hand pane, the velocity magnitude is shown at an early stage in the collision at $t = 0.8$\,Myr after the beginning of the simulation. The gas filaments formed by the turbulence lie at a range of angles to the collision axis, creating a shocked interface at the cloud contact point that is not straight, but follows these oblique angles. As with the thin disk previously seen in section~\ref{sec:results_noturb}, ripples are formed along the shocked filaments with a spatial scale much smaller than the filament length. These ripples are then enhanced by the thin shell instability. 

As {\it Cloud 1} penetrates the larger {\it Cloud 2}, the oblique arc of the overall shock front focussed on the concave parts of the shocked filaments and increases their mass and density. These sections begin to gain inertia, causing them to slow with respect to the shock front and creating a displacement that increases with time. This can be seen in the right-pane of Figure~\ref{fig:shockfront}.

\subsection{1D density distributions}

\begin{figure*}
\begin{center}
\includegraphics[width=1.0\textwidth]{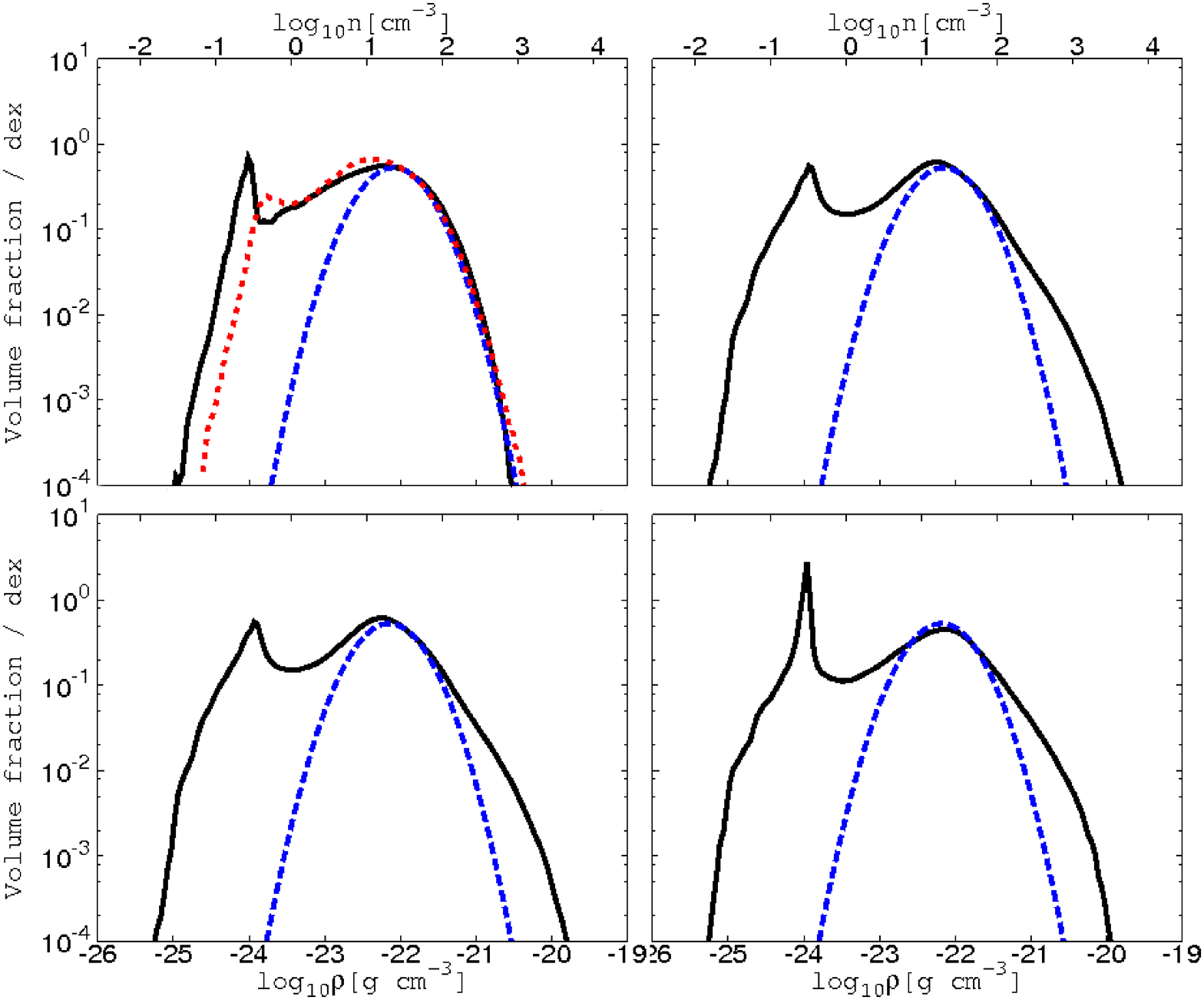}
\caption{Probability distribution functions (PDF) for the cloud collisions at different relative velocities. The blue dashed line shows a log-normal fit. Top-left shows the comparison static case for the gas in {\it Cloud 2} when it does not undergo a collision. The black solid line depicts the profile at 1.0\,Myr while the red dotted line is the gas at $t_{ff_1} = 5.3$\,Myr. The high density gas maintains a log normal fit over the duration of the simulation. Top-right shows the result for the 3.0\,km/s collision case, bottom-left the 5.0\,km/s and right the 10.0\,km/s simulation. In all three cases, the black solid line shows the profile when the maximum number of cores are present. Upon forming the cores, the gas deviates from the log-normal to grow an extended tail.}
\label{fig:pdfs}
\end{center}
\end{figure*}

After the initial formation of the oblique shock front, the colliding clouds take on visually different forms, depending on their relative collision velocity. In the middle and final panes of Figure~\ref{fig:collvel}, the combined gas from the colliding clouds forms an open arc that is marked more strongly in the higher velocity collisions. This distinctive shape resembles the observations performed by Torii et al., in prep. of the Galactic HII region, RCW120, which they postulate is the result of a cloud collision. At the final time shown, dense structures can be seen in the 3.0\,km/s and 5.0\,km/s case, but gas remains at a lower density in the 10\,km/s simulation. 

The overall gas density of the clouds can be explored in the one-dimensional probability distribution functions (PDF) which are shown in Figure~\ref{fig:pdfs} for the non-collision case and the three different collision velocities considered. Gas is included in the PDF if it is within a 7\,pc radius sphere centred on {\it Cloud 2}. The blue dashed line in all panes marks a log-normal profile with equation:

\begin{equation}
f(x;\mu,\sigma) = \frac{A}{\sigma \sqrt{2\pi}}e^{-\frac{1}{2}\left( \frac{x-\mu}{\sigma} \right)^2},\label{eq:Gauss}
\end{equation}

\noindent where $x = \ln \rho/\bar{\rho}$ and the constants have values $A \sim 1.2$, $\mu \sim -0.3$ and $\sigma = 0.9$.

In the top-left pane of Figure~\ref{fig:pdfs}, the static evolution of {\it Cloud 2} is shown, without any collision event. The black solid line shows the gas at 1.0\,Myr and the red dotted line shows the profile after the free-fall time of the smaller {\it Cloud 1}, $t_{\rm ff_1} = 5.3$\,Myr; the duration of that simulation and longer than the shock crossing time in either the 5.0\,km/s or 10\,km/s case. In the static cloud simulation, the gas above a density of 10.0\,cm$^{-3}$ closely follows a log-normal distribution throughout the simulation. At lower densities, background gas from around the cloud forms the peak at $\sim 0.5$\,cm$^{-3}$.

However, the next three plots show clear deviations in that higher density region. Top-right shows the result for the 3.0\,km/s case while on the bottom row the left-hand pane shows the 5.0\,km/s simulation and right-hand pane is for the 10.0\,km/s simulation. The times shown correspond to when the maximum number of cores ($\rho_{\rm thresh} = 10^{-20}$\,gcm$^{-3}$) occurs during the collision; 4.6\,Myr, 3.7\,Myr and 2.2\,Myr for collision times 3, 5 and 10\,km/s respectively. All three collisions cause a tail to develop in the high density gas, moving the profile away from a single log-normal. This extension is most marked in the 3.0\,km/s and 5.0\,km/s collisions and is present, but less extended, in the 10.0\,km/s run. We will see in section~\ref{sec:core_number} that the tail corresponds to the formation of the potentially star-forming cores and in the 10.0\,km/s simulation, the collision is completed (shock wave exited {\it Cloud 2}) before a gravitationally bound core can be formed. 

This result --that deviation from the log-normal is linked with star formation-- agrees strongly with the observations from \citet{Kainulainen2009}, who find that giant molecular clouds without active star formation show a log-normal distribution, while clouds displaying star formation show prominent non-log-normal wings. This result was also found in simulations by \citet{Federrath2013} who modelled the evolution of a periodic box of isothermal star forming gas under a variety of different conditions for turbulent driving and magnetic fields. They found in all cases where the star formation efficiency was not zero, the PDF developed a tail. 

A related result was found by \citet{Tremblin2014}, who found a double PDF structure in clouds containing an expanding HII region. They concluded that this was a signature of triggered star formation, with an additional power-law tail from the gravitational effects. Such a double PDF structure would support the extended tail in Figure~\ref{fig:pdfs}, although the second peak is not strongly visible.

\subsection{Turbulence characterisation}
\label{sec:b_parameter}

\begin{figure}
\begin{center}
\includegraphics[width=1.0\columnwidth]{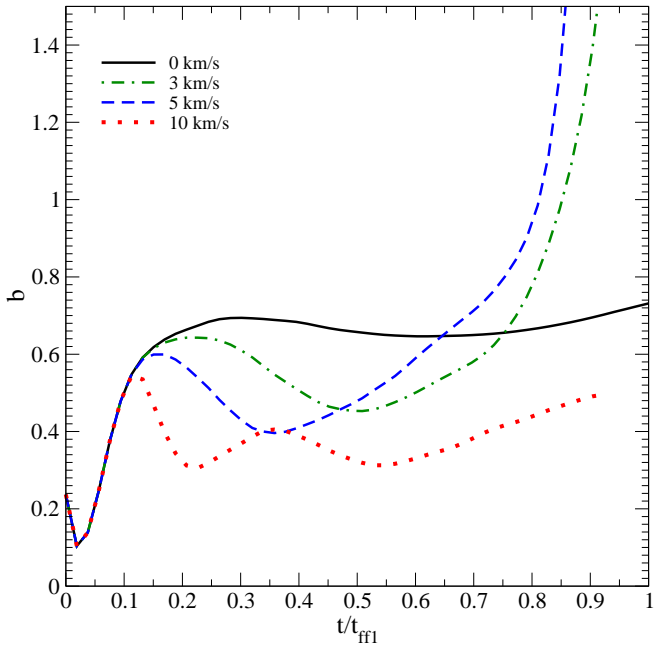}
\caption{Characterisation of the turbulence during the simulation in {\it Cloud 2} for the three runs at varying collision speeds. A $b$-parameter around $1/3$ suggests solenoidal driving of the turbulence whereas $b \sim 1$ implies compressive turbulence dominates. For the non-static cases, the first peak marks the start of the collision, which causes a reduction in $b$. As gravity takes control to form the cores, the $b$ value escalates}
\label{fig:b}
\end{center}
\end{figure}

The shape of the probability distribution functions can be directly related to the type of turbulence inside the cloud. For an isothermal gas, the width of the PDF grows with increasing Mach number, where the value of the proportionality constant is determined by the modes of turbulence present \citep{Federrath2008, Passot1998}. Since the majority of the cloud gas sits close to the bottom end of our cooling curve at 10\,K, we can apply the relation for a non-magnetised gas between these three quantities \citep{Burkhart2012, Padoan1997}:

\begin{equation}
\sigma_s^2 = \ln (1+b^2\mathcal{M}^2)
\label{eq:b}
\end{equation}

\noindent where the width of the PDF is determined by the logarithmic standard deviation, $\sigma_s$, in $s \equiv \ln (\rho/\bar{\rho})$, the natural logarithm of the density divided by the mean density. The cloud Mach number is defined as $\mathcal{M} \equiv <\sigma_{3D}/c_s>$ for speed of sound, $c_s$, and three-dimensional velocity dispersion, $\sigma_{3D}$.

In simulations performed by \citet{Federrath2008}, the value of the proportionality term, $b$, was found to be dictated by how the turbulence was being driven. When compressive forcing was used in their calculations, the standard deviation was three times larger than with solenoidal (divergence-free) forcing. The former gave a value of $b \sim 1$ while solenoidal turbulence produced $b \sim 1/3$.

Rearranging equation~\ref{eq:b} for $b$ gives $b = \sqrt{(\exp(\sigma_s^2)-1)/\\ \mathcal{M}^2}$, which is plotted in Figure~\ref{fig:b} for {\it Cloud 2} over the duration of the simulation. As with the PDF plots in Figure~\ref{fig:pdfs}, gas is included in the calculation for $b$ if it is within a 7\,pc radius sphere centred on {\it Cloud 2}. The solid black line shows the static evolution of {\it Cloud 2} in the absence of any collision, while the green dot-dash line shows the results for the 3\,km/s collision, the blue-dashed line marks the 5\,km/s evolution and the red dotted line is for the 10\,km/s run. In all cases, the turbulence is injected into the equilibrated sphere at $t/t_{ff_1} = 0$, which begins to cool, lowering the sound speed and causing $b$ to drop. As the turbulence begins to decay, the velocity dispersion decreases and $b$ rises. At the point where {\it Cloud 1} begins to move and the internal density structure of both clouds has reached the lognormal profile seen in Figure~\ref{fig:pdfs}, $t/t_{ff_1} = 0.1$ and $b \sim 0.5$, suggesting a mix of compressional and solenoidal modes in the filamentary structure created by the turbulence.

In the absence of a collision, the Mach number and density standard deviation tend to roughly constant values to form the weakly filamentary structure shown in the right-hand panel of Figure~\ref{fig:image-noturb-turb}. This corresponds to a steady $b$ value of around 0.7. In the case of a collision, the velocity dispersion and Mach number initially rise as {\it Cloud 1}'s material meets {\it Cloud 2}. This first contact is prior to the shock forming, so the standard deviation in density remains unchanged, causing the value of $b$ to drop. As {\it Cloud 1} penetrates {\it Cloud 2}, the shock front forms with an accompanying jump in the density deviation. This occurs at $t/t_{ff_1} \simeq 0.38, 0.28$ and $0.19$ for the 3\,km/s, 5\,km/s and 10\,km/s case respectively. The density within the shock front continues to grow as {\it Cloud 1} pushes forward and the $b$ value turns and begins to increase, driving into the compressional regime. This turning point occurs most rapidly after the clouds touch in the 10\,km/s case, since the stronger shock raises the density standard deviation most quickly. In the case of the 3\,km/s and 5\,km/s runs, the density standard deviation continues to grow and $b$ escalates. We will see in the next section that this rapid increase in $b$ corresponds to the formation of a gravitationally bound core, whose mass undergoes runaway growth. The 10\,km/s collision begins this process, but the shock wave exits the cloud before the density standard deviation gets very high. This rapid progression will actually turn out to prevent a bound core from forming. 

As described in the previous section, the shape of the density PDF is associated both from observations and simulations with the cloud's star formation. Since the width of the PDF depends on $b$, this association can be carried through to explore the link between star formation and turbulence type. In their paper, \citet{Federrath2012} derived the $b$ dependence from six analytical star formation models that are based on the density PDF. They found that $b$ positively correlated with star formation efficiency, with they ascribed to the compressional modes being more effective at forming more dense star-forming regions. Our results support this conclusion, with the formation of the shock front corresponding to a marked rise in $b$ that takes it into the compressional regime. This then forms cores which become gravitationally bound and collapse, causing self-gravity to dominate the density field over turbulence and escalating the value for $b$.

\subsection{Core formation}
\label{sec:core_formation}

As described in section~\ref{sec:setup_cores}, gas which has the potential for star formation is defined as a `core' when it reaches a threshold density of $\rho_{\rm thresh} = 10^{-20}$\,gcm$^{-3}$. Using this definition and the tracking described in the same section, we identified the cores in the three simulations and explored their properties. 

\subsubsection{Core number evolution}
\label{sec:core_number}

\begin{figure*}[!thb]
\begin{center}
\includegraphics[width=1.0\textwidth]{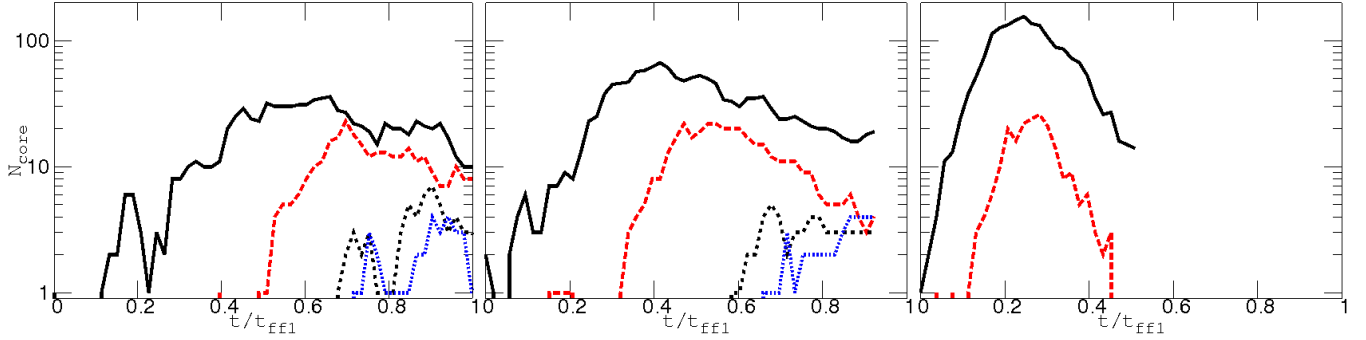}
\caption{Core number evolution as a fraction of {\it Cloud 1}'s free-fall time for simulations at three different collision velocities. From left to right, the panels show the 3\,km/s run, 5\,km/s run and 10\,km/s run. $t = 0$ corresponds to the time when the clouds are just touching. The lines represent cores formed at different density thresholds. The black solid line is for a core density of $5\times 10^{-21}$\,gcm$^{-3}$. The red dashed line is our usual definition of $1\times 10^{-20}$\,gcm$^{-3}$. The dot-line marks a threshold of $5\times 10^{-20}$\,gcm$^{-3}$ and the blue chain line is at $1\times 10^{-19}$\,gcm$^{-3}$. As the collision velocity increases, the lower density cores (black solid line and red dashed) increase in number as the cores form faster. However, in the 10\,km/s simulation, the higher density cores do not have time to form before the shock has passed through the cloud.}
\label{fig:clumpnum}
\end{center}
\end{figure*}

Unlike when turbulence was not included in section~\ref{sec:results_noturb}, multiple cores with densities higher than our core threshold, $\rho_{\rm thresh} = 10^{-20}\,{\rm g cm^{-3}}$, are formed in all three runs performed at different collisional velocities. The evolution of the core number plotted as a fraction of {\it Cloud 1}'s free-fall time is shown in Figure~\ref{fig:clumpnum} until 1.0\,Myr after the shock exits {\it Cloud 2}. Left to right show the core evolution for the runs with collisional velocity 3\,km/s, 5\,km/s and 10\,km/s. The initial time in the plots, $t = 0$, corresponds to the clouds just prior to collision as their surfaces touch. Each plot considers four different density values for the core definition to provide a quantitative feel of the fragmentation of the gas: $\rho = 5\times 10^{-21}\,{\rm g cm^{-3}}$ (black solid line), $\rho_{\rm thresh} = 10^{-20}\,{\rm g cm^{-3}}$ (red dashed line), $\rho = 5\times 10^{-20}\,{\rm g cm^{-3}}$ (dot line) and $\rho = 10^{-19}\,{\rm g cm^{-3}}$ (blue chain line). As mentioned in the earlier sections, observationally star formation is seen to occur at densities around $10^4$\,cm$^{-3} \approx 3\times 10^{-20}$\,gcm$^{-3}$ \citep{Lada2010}. 

In all simulations, core formation at our main threshold density (red dashed) and at the lower threshold value (black solid) begins at a time much shorter than the free-fall time of {\it Cloud 1} or {\it Cloud 2} ($t_{\rm ff_2} \simeq 1.4t_{\rm ff_1}$), ruling out the possibility of core production purely from the cloud's natural gravitational collapse. As the clouds touch and the shock front first appears, a small number of short-lived cores are formed at our lowest threshold. As {\it Cloud 1} penetrates {\it Cloud 2}, the bow shock develops and presses on the cloud interface, destroying this first population. A second generation of cores develops in the newly shaped shock which gain in density to form the higher threshold populations.

The time for the first core formation and the peak in the maximum number of cores is dependent on collision velocity, with a higher relative speed creating cores more rapidly. For our threshold density, cores begin to form at 0.39\,$t/t_{ff_1}$ in the 3\,km/s simulation which is reduced to $0.15$ in the 5\,km/s collision and $0.03$ in the 10\,km/s collision. The time for the maximum core number reduces from $0.65, 0.41$ to $0.24$ as we increase collision speed. In the case of the two lower core thresholds, the increase in velocity also corresponds to a higher maximum core count, with three times as many cores formed at the lowest threshold in the 10\,km/s run compared with the 3\,km/s case. The ability to locally collapse gas into a core is therefore strongly dependent on the shock-front strength, with stronger shocks making it easier to create a local gravitational collapse.

As the shock leaves {\it Cloud 1}, the post shock gas begins to expand and drop in density. This causes the core number to begin to lower as non-bound cores dissipate. The rate of dissipation is also related to shock speed, with the faster collision resulting in gas having less time to grow in density within the shock front, leading to a more rapid reduction in core number once the shock has past. In the 3\,km/s and 5\,km/s simulation, one bound core is formed that persists through to the end of the simulation. 

The need for the shock front to remain in the cloud long enough to allow core growth is seen more clearly when the core threshold density is increased to consider the formation of objects at $\rho = 5\times 10^{-20}\,{\rm g cm^{-3}}$ and $\rho = 10^{-19}\,{\rm g cm^{-3}}$, where the previous trends appear to break down when we reach the 10\,km/s collision. In both the 3\,km/s and 5\,km/s cases, higher density cores are formed after $t/t_{ff_1} = 0.6$, with the formation occurring earlier for the 5\,km/s simulation in the same pattern as for the lower density cores. However, in the 10\,km/s simulation, no cores are seen after $t/t_{ff_1} = 0.5$. At this stage, the faster shock generated in the 10\,km/s case has passed through {\it Cloud 1} entirely, before the gas has had the chance to reach the higher density thresholds. With none of the cores formed becoming gravitationally bound, the post-shock gas swiftly drops in density, leaving no gas above any of the considered thresholds.

\subsubsection{Core properties}
\label{sec:core_properties}

\begin{table}[htdp]
\begin{flushleft}
\newlength{\myheight}
\setlength{\myheight}{0.5cm}
\begin{tabular}{cccccc} \hline
\parbox[c][\myheight][c]{0cm}{}M[$M_{\odot}$] & $r_{\rm core}$[pc] & $\bar{n}$[cm$^{-3}$] & $\sigma$[km s$^{-1}$] & $\alpha_{\rm vir}$ &  $\dot{M}$[$M_{\odot}$yr$^{-1}$]  \\ \hline
47.85 & 0.35 & 2.55E+04 & 0.34 & 1.52 & 4.23E-04\\ 
14.04 & 0.24 & 1.17E+04 & 0.21 & 2.07 & 3.42E-04\\ 
4.35 & 0.18 & 5.93E+03 & 0.13 & 3.59 & 3.40E-05\\ 
3.06 & 0.16 & 5.80E+03 & 0.20 & 6.23 & 2.28E-05\\ 
2.21 & 0.15 & 4.14E+03 & 0.18 & 7.34 & 1.18E-05\\ 
 \hline
\end{tabular}
\end{flushleft}
\caption{Properties of the five most massive cores at the point of maximum core number, 3.5Myr, in the 3CC simulation}
\label{table:cc3}
\end{table}

\begin{table}[htdp]
\begin{flushleft}
\newlength{\myheighta}
\setlength{\myheighta}{0.5cm}
\begin{tabular}{cccccc} \hline
\parbox[c][\myheighta][c]{0cm}{}M[$M_{\odot}$] & $r_{\rm core}$[pc] & $\bar{n}$[cm$^{-3}$] & $\sigma$[km s$^{-1}$] & $\alpha_{\rm vir}$ &  $\dot{M}$[$M_{\odot}$yr$^{-1}$]  \\ \hline
18.15 & 0.26 & 1.16E+04 & 0.27 & 2.31 & 2.11E-04\\ 
6.83 & 0.20 & 7.00E+03 & 0.17 & 3.09 & 1.08E-04\\ 
3.90 & 0.17 & 7.27E+03 & 0.19 & 4.94 & 2.55E-05\\ 
2.85 & 0.15 & 7.81E+03 & 0.17 & 5.60 & 1.37E-05\\ 
1.79 & 0.13 & 6.80E+03 & 0.11 & 5.71 & 1.06E-05\\   
 \hline
\end{tabular}
\end{flushleft}
\caption{Properties of the five most massive cores at the point of maximum core number, $2.5$\,Myr in the 5CC simulation}
\label{table:cc5}
\end{table}

\begin{table}[htdp]
\begin{flushleft}
\newlength{\myheightb}
\setlength{\myheightb}{0.5cm}
\begin{tabular}{cccccc} \hline
\parbox[c][\myheightb][c]{0cm}{}M[$M_{\odot}$] & $r_{\rm core}$[pc] & $\bar{n}$[cm$^{-3}$] & $\sigma$[km s$^{-1}$] & $\alpha_{\rm vir}$ &  $\dot{M}$[$M_{\odot}$yr$^{-1}$]  \\ \hline
0.50 & 0.09 & 4.06E+03 & 0.22 & 24.11 & 6.18E-06\\
0.48 & 0.09 & 4.51E+03 & 0.30 & 33.17 & 4.99E-06\\
0.46 & 0.08 & 4.52E+03 & 0.22 & 23.59 & 7.12E-06\\
0.45 & 0.09 & 4.40E+03 & 0.27 & 31.80 & 5.39E-06\\
0.44 & 0.08 & 4.83E+03 & 0.43 & 58.34 & 7.72E-06\\
 \hline
\end{tabular}
\end{flushleft}
\caption{Properties of the five most massive cores at the point of maximum core number, $1.4$\,Myr in the 10CC simulation}
\label{table:cc10}
\end{table}

\begin{figure*}[!t]
\begin{center}
\includegraphics[width=1.0\textwidth]{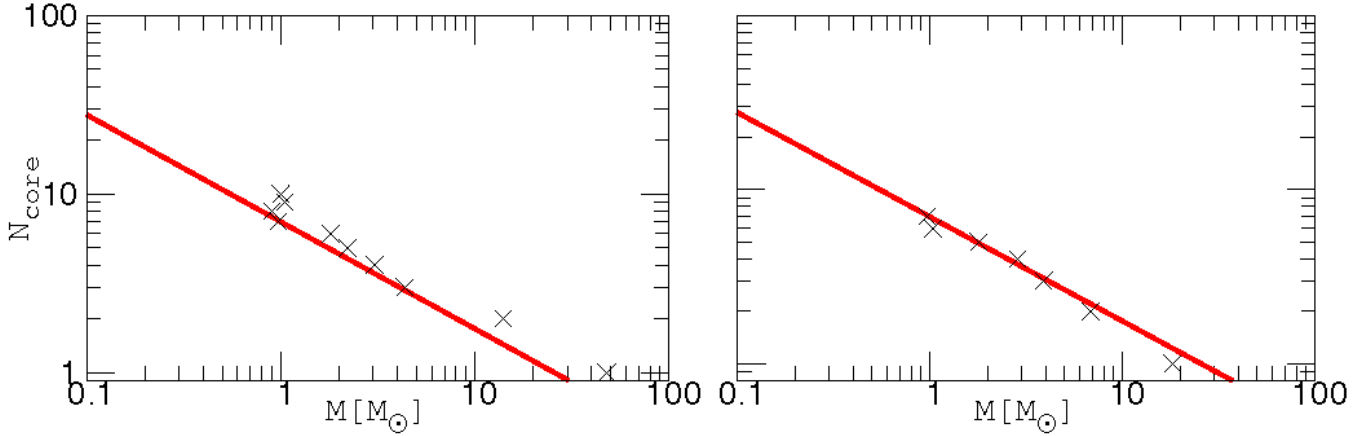}
\caption{Cumulative Mass Distribution at the time of maximum core number. Plots show result from 3CC (left) and 5CC (right). The black crosses mark the mass of the cores while the red solid line shows the fit $N_{\rm core} = 7M^{-0.6}$, giving a value of $\gamma = 1.6$ for Equation~\ref{eq:cmd} in agreement with the observed value for the Orion molecular cloud.}
\label{fig:CMF}
\end{center}
\end{figure*}

The properties of each of the cores formed in the simulation are shown in Tables~\ref{table:cc3}, \ref{table:cc5} and \ref{table:cc10} where we list the most massive five cores. For each core found, the mass, radius, average number density, velocity dispersion, virial parameter and estimated accretion rate are shown. The core radius is defined from the core volume, $V_{\rm core}$, as $r_{\rm core} =  \left( \frac{3 V_{\rm core}}{4 \pi} \right)^{1/3}$. The virial parameter, $\alpha_{\rm vir}$, is a measure of the ratio of the kinetic and potential energy for a spherical object, defined by: 

\begin{equation}
\alpha_{\rm vir} = \frac{5\sigma_{\rm th}^2 r_{\rm core}}{GM}
\end{equation}

\noindent with $\sigma_{\rm th}$, the 1D velocity dispersion of the core, including the sound speed: $\sigma_{\rm th} \equiv \left(c_s^2 + \sigma^2_{\rm 1D}\right)^{1/2}$. A value $\alpha_{\rm vir} < 1$ implies the core is virialised with the total kinetic energy of the cloud equalling half its gravitational. A gravitationally bound core should correspond to an $\alpha_{\rm vir} < 2.0$.

Comparing the core properties across the three different collisional velocities, the effect of the shock wave's duration inside the cloud is once again apparent; the slower the collision, the longer the gas spends inside the shock front, allowing a higher mass core to be formed. So while the number of cores increases with higher collision speed, the core mass obtained is related to the length of time the shock remains inside the cloud. This results in the most massive core forming in the 3CC simulation while the 10CC cores remain small, despite their numerously seen in Figure~\ref{fig:clumpnum}. The average density of the 10CC cores hovers at our threshold definition around $10^4$\,cm$^-3$ with the average core accretion rate over a factor of 10 lower than in the 3CC and 5CC case due to the fast exit of the shock front. This result suggests that maximum core size is determined by accretion, not by the initial fragmentation. We will revisit this topic in Section~\ref{sec:core_growth}.

Cores are typically unbound, with only the largest core in simulation 3CC and 5CC obtaining an $\alpha_{\rm vir}$ close to the threshold for gravitational binding. In fact, a more accurate calculation of this property reveals both those cores are bound, but the rest remain unbound. This is reflected in their mass, with the bound cores reaching a mass several times larger than the next biggest candidate through a runaway growth. This mass increase will also be explored in Section~\ref{sec:core_growth}. The cores continue to grow past the time shown in the tables, but the unbound cores ultimately begin to lose mass once the shock has passed and dissipate as shown in the core number plots in Figure~\ref{fig:clumpnum}. The number of unbound cores makes it possible that only a single object in these small clouds will survive to collapse and form a massive star or super star cluster.

\begin{figure*}[!t]
\centering
\includegraphics[width=\textwidth]{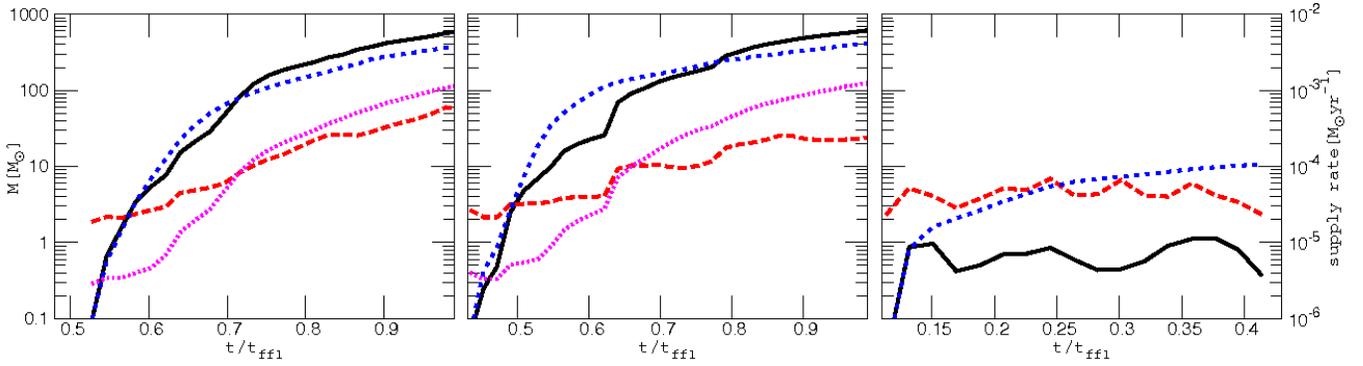}
\caption{Mass evolution of the bound core. Left to right shows the core formed in simulations 3CC, 5CC and 10CC. Since no bound core is formed in the 10CC case, the evolution of the most massive core is shown. The black line marks the core mass, the blue dotted line is the accreted mass, the red dashed line is the effective Jeans mass and the pink dashed line shows the supply rate; the ratio of the core's collapsing mass to its free-fall time, ${\rm max}(0.5M_{\rm core},M_{\rm J,eff})/ t_{\rm ff}$ whose axis is on the right.}
\label{fig:BM}
\end{figure*}

The core masses can be explored by looking at the cumulative mass distribution (CMD) shown in Figure~\ref{fig:CMF}. Here, we restrict the cores plotted to those only containing a minimum of 27 cells (approximately 3 cells in each dimension), to ensure the best resolution. Due to the size of the cores formed in the 10CC case, the cores were not resolved enough to plot the CMD for this run. The CMD is derived from the core mass function, $\phi$, defined as:

\begin{equation}
\phi \equiv \frac{dN}{dM} \propto M^{-\gamma}
\label{eq:cmd}
\end{equation}

\noindent where $M$ is the core mass and $dN$ is the number of cores with masses between $M$ and $M+dM$. The cumulative mass function is then given by the integral:

\begin{equation}
N(>M) = \int^{\infty}_{M} \frac{dN}{dM} dM \propto M^{-(\gamma-1)}
\end{equation}

The red solid line in Figure~\ref{fig:CMF} shows a power-law with equation $N_{\rm core} (>M)= 7M^{-0.6}$, giving a $\gamma = 1.6$. This value agrees with observational results from \citet{Tatematsu1993}, who found cores in the Orion A molecular cloud fit a value of $\gamma = 1.6 \pm 0.3$ for $M \gtrsim 50 M_{\odot}$ (the completeness limit of their survey). At masses closer to our own cores, \citet{Scalo1985} found a $\gamma = -1.5$ to $-1.0$ for Orion A cores with $M \ge 4$\,$_\odot$ and \citet{Stutzki1990} found $\gamma = -1.7$ for $M \ge 10$\,M$_\odot$ for cores in the M17 SW region. In Orion B, \citet{Johnstone2001} found a $\gamma =-1.5-2.0$ for $M > 1.0 $\,\msun. In general, these values are similar to the Salpeter stellar IMF \citep{Salpeter1955}, although it remains unclear whether the core and ultimate stellar size are directly connected. Simulations of the collapse of a single isothermal cloud performed by \citet{Klessen1998} also found agreement with $\gamma = -1.5$ for cores formed in their fragmented gas, while a comparison of clouds fragmenting with different initial density profiles by \citet{Girichidis2011} likewise found a Salpeter consistent value of $\gamma = -1.35$ for all models. These similarities suggest that triggered core formation should take a similar form to more quiescent core creation. However, the time-scale is liable to be significantly shorter since even in the absence of turbulence, the free-fall time of the clouds is longer than their collisional interaction. This potentially produces a different IMF, as the fast-forming cores may be able to gain a higher mass before feedback effects stunt their growth.

\subsubsection{Bound core evolution}
\label{sec:core_growth}

From the point of star formation, the most interesting structures to develop will be the gravitationally bound cores. These objects are the ones that are the most likely to collapse and form stars. As shown in the previous sections, despite many cores being produced in the collision, only a single bound core is formed in the 3CC and 5CC runs. Due to the collision process being short in the 10\,km/s case, this run does not form any bound cores. 

A bound core can potentially form in two ways: it can collapse directly from shocked gas to form a sufficiently dense structure that its potential energy already exceeds its kinetic. Alternatively, it can begin life as a unbound core and gather mass via accretion of the surrounding diffuse gas or through a merger with another core. The evolution of the bound cores is shown in Figure~\ref{fig:BM}, plotted over the fraction of {\it Cloud 1}'s free-fall time. The black line marks the core's actual mass while the red dashed line marks the core's effective Jeans mass as defined by $M_{\rm J,eff} = \sigma_{\rm th}^{3}/(6.0 G^{3/2}\rho^{1/2})$. The blue dotted line shows the expected accreted mass gain, with the accretion rate defined as:

\begin{equation}
\dot{M} = \pi r_{\rm acc}^2\sigma_{\rm th, acc}\rho_{\rm acc}
\end{equation}

\noindent where $\sigma_{\rm th, acc}$ and $\rho_{\rm acc}$ are the average thermal velocity dispersion and density of a sphere surrounding (but not including) the core with radius, $r_{\rm acc}$, given by the Bondi radius:

\begin{equation}
r_{\rm acc} = \frac{2 G M}{\sigma_{\rm th}^2} + r_{\rm core}
\end{equation}

\noindent upto a maximum of $1$\,pc, the width of the shock front \citep{Bondi1951}. Note that since we do not include the core gas in the accretion calculation, we add the core radius to the Bondi radius to ensure a reasonable sample of gas outside the core. Finally, the pink dashed line shows the gas supply rate, defined as the half the core's mass or its effective Jeans mass (whichever is larger) over its free-fall time, ${\rm max}(0.5M_{\rm core}, M_{\rm J,eff})/t_{\rm ff}$. 

Left to right in Figure~\ref{fig:BM} are the results for the 3CC, 5CC and 10CC runs. Since no bound core is formed in the 10CC case, the mass evolution of the largest formed core is plotted.  As this is unlikely to collapse, we do not plot the supply rate in this case. 

In Figure~\ref{fig:clumpnum}, the first cores to form during the collision are low density, with the high density cores only appearing once the gas has spent time in the shock front. This suggests that cores do not initially form bound, but increase their mass over time. What cannot be seen so easily from Figure~\ref{fig:clumpnum}, is whether this mass gain occurs through mergers or accretion. 

\begin{figure*}
\begin{center}
\includegraphics[width=1.0\textwidth]{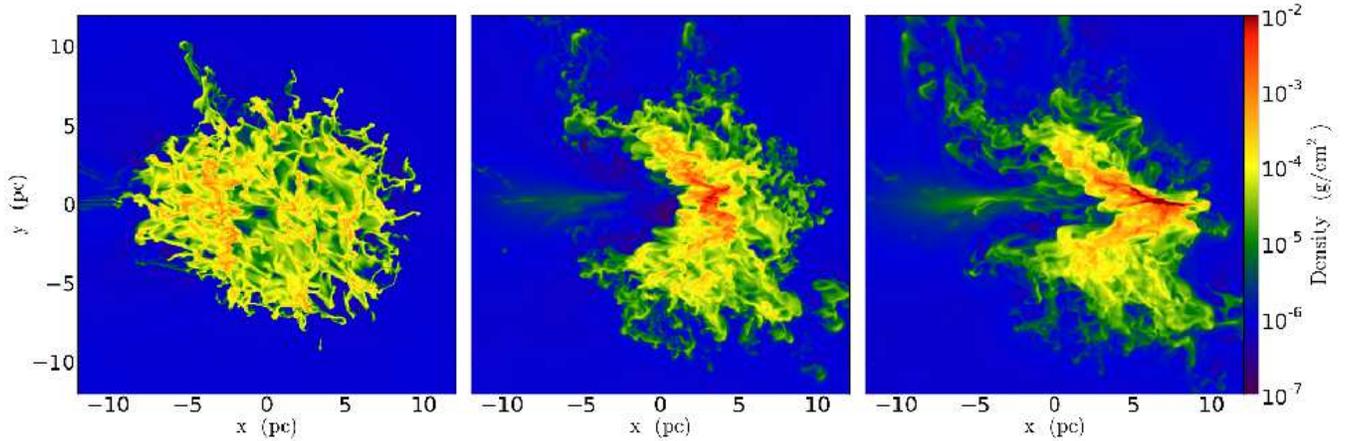}
\caption{Surface density evolution for the high resolution 5CCHR simulation. Images times and properties match those of the middle panel in Figure~\ref{fig:collvel}. The evolution mirrors the 5CC run, with the characteristic arc of shocked gas fragmenting to form cores.}
\label{fig:Ref3_images}
\end{center}
\end{figure*}

In the 3\,km/s simulation, the core growth is predominantly by accretion. It surpasses the effective Jeans mass at $t/t_{\rm ff_1} \sim 0.55$ and grows to a mass of 700\,\msun. The same is almost true for the bound core in the 5\,km/s simulation. Growth is predominantly by accretion, yet a jump can be seen between $0.6-0.7$ that represents a merger scenario. It is therefore likely that cores grow by both accretion and mergers with other neighbouring cores, with accretion playing the most dominant role. In the 10\,km/s simulation, the core growth is insufficient to be gravitationally bound before $t/t_{\rm ff_1} = 0.3$, the time for the shock wave to propagate through {\it Cloud 1}. The accretion rate can initially explain the early core growth until  $t/t_{\rm ff_1} = 0.15$, after which the core struggles to gain mass. This may be due to the exit of the shock from the core's region, which causes the hydrodynamic flow around the low mass core to disrupt it. 

In reality, the maximum mass of the star forming from a gas core will be limited by the radiative heating from the forming stellar object. Once the star has formed, Bondi-Hoyle accretion (as plotted by our blue dashed line in Figure~\ref{fig:BM}) is expected to only be efficient up to masses of 10\,\msun. From Figure~\ref{fig:BM}, the bound cores in the 3CC and 5CC simulations reach their Jeans mass at approximately $2$\,\msun and $3$\,\msun respectively. If stars are formed at this point, $1-2$\,\msun stars will be created, allowing for 50\% of the core's gas to escape as outflows as suggested by the analytical estimates of \citet{McKee2002, Matzner2000} and the simulation comparisons of \citet{Federrath2012}. 

However, if the cores are prevented from collapsing (for instance by magnetic fields or turbulence unresolved in our simulation) then the core may continue to follow the black line in Figure~\ref{fig:BM} before forming a star. In this situation, the core will only collapse to form a massive star if the ram pressure from the infalling gas can exceed that of the outward radiation pressure. Using dimensional arguments, \citet{McKee2002} argue that the rate of gas flow during the collapse into a star should be given by $\dot{m}_\star \simeq \frac{m_*}{t_{\rm ff}}$, where $m_*$ is the instantaneous stellar mass. Using this, we can argue that $m_*$ should be either half the core's mass or the effective Jeans mass (since gas mass below this value will not collapse), whichever is larger. This gives the rate of gas feeding the new star as $\dot{m}_\star \simeq \frac{{\rm max}(0.5M_{\rm core},M_{\rm eff, J})}{t_{\rm ff}}$, which we plot as the pink dashed line in Figure~\ref{fig:BM}, with the axis shown on the right. If the core is able to grow to 200\,\msun ($m_* = 100$\,\msun), then the supply rate will be $3\times 10^{-4}$\,\msun yr$^{-1}$, a value very close to the $6\times 10^{-4}$\,\msun yr$^{-1}$ rate that \citet{McKee2002} estimate is required to form a 100\,\msun star.

\section{High resolution study}
\label{sec:highres}

\begin{figure*}
\begin{center}
\includegraphics[width=1.0\textwidth]{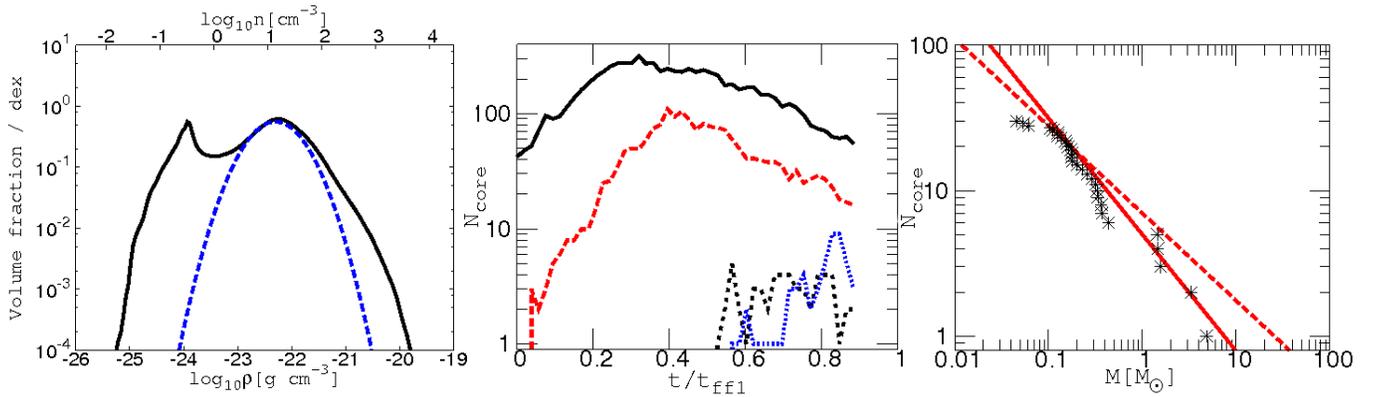}
\caption{
The PDF, core number evolution and core mass distribution for the high resolution 5CCHR simulation. The results are similar to the lower resolution 5CC counterpart in Figures~\ref{fig:pdfs}, \ref{fig:clumpnum} and \ref{fig:CMF}, with the main differences being abundance and earlier appearance of low mass cores. In the CMD (right), the red solid line shows the fitted power-law with $\gamma = -1.8$, while the dashed line shows the original $\gamma = -1.6$ fit for the 5CC simulation.}
\label{fig:Ref3_plots}
\end{center}
\end{figure*}

To assess how our results are affected by resolution, we reran the 5\,km/s relative velocity collision case with a limiting resolution of 0.03\,pc, increasing the resolution from our model in section~\ref{sec:results} by a factor of 2. The evolution of the collision is shown in the surface density images in Figure~\ref{fig:Ref3_images} for the same times as for the 5CC case in the middle row of Figure~\ref{fig:collvel}. The progression of the collision is very similar in both the low and high resolution runs, with the smaller cloud penetration producing the characteristic arc of shocked gas which then collapses into a dense region of cores.

The quantitative study of the gas properties are shown in Figure~\ref{fig:Ref3_plots} which shows the gas PDF (left), core number evolution (center) and the core cumulative mass distribution (right). The gas PDF is almost identical to its low resolution counterpart in Figure~\ref{fig:pdfs}, with the high density gas formed a non-log-normal wing. This suggests the quantity of core-forming gas is the same in the higher resolution run, but the core number evolution (middle pane) shows that the extra refinement does allow a larger number of cores to be formed at the lower two threshold densities. In the case of our lowest threshold, $\rho = 5\times 10^{-21}$\,gcm$^{-3}$, cores are formed before the collision, but the higher threshold cores remain forming only within the collisional shock. The ability to follow the collapse of smaller objects also allows cores to be formed earlier, allowing them potentially more time to accrete within the shock front. This added bonus, however, only has a small benefit to the higher mass cores who appear at about the same time in both the high and low resolution runs at $t/t_{\rm ff_1} \simeq 0.6$. 1\,Myr after the shock exits {\it Cloud 2} (the cut-off point for the core number evolution plot), both the 5CC and 5CCHR have formed 1 bound core, with this number increasing to 3 bound cores for 5CCHR 0.5\,Myr later.  

The right-hand pane of Figure~\ref{fig:Ref3_plots} shows the cumulative mass distribution in the higher resolution case. The solid red line shows the fit $N_{\rm core} (>M)= 5M^{-0.8}$, giving a $\gamma = -1.8$. This is very close (though slightly steeper) than the $\gamma = -1.6$ 5CC fit, which is shown as a red dashed line. The added refinement has pulled in the maximum mass the cores achieve to below 10\,M$_\odot$, while increasing the number of low mass cores below $M < 1$\,M$_\odot$. 

Overall, the evolution of the higher resolution case is extremely close to that of its low resolution counterpart. The same overall quantity of dense gas is formed in the shock which translates to a similar number of high density and bound cores. 

\section{Discussion and Conclusions}

We explored the formation and evolution of pre-stellar gas cores in a collision of non-identical clouds with Bonner-Ebert profiles using hydrodynamical simulations. For the majority of our runs, the limiting resolution was 0.06\,pc, with one additional run performed at 0.03\,pc. We compared the effect of collisional speed on the core formation and examined the development of bound cores formed during the impact. 

In all cases, the collision between the two clouds produces a shocked disc of gas where the clouds' surfaces meet. Due to the size difference between the two clouds, the shock front becomes oblique as the clouds merge, pressing down on the shocked disc to form a crescent structure that is commonly observed in observations of suspected cloud-cloud collision events \citep{Ohama2010, Furukawa2009}.  

Without turbulence, the smaller cloud is then compressed by its own gravity and the oblique shock, creating a single core. When turbulence is added, the shock front is comprised of filaments which lie at a range of angles to the propagating wave. These form thin shell instabilities, producing ripples that ultimately fragment into multiple cores.

The number of these cores produced is proportional to the relative collision speed between the clouds. A faster collision forms cores more quickly and in greater numbers. However, the subsequent growth of these cores is controlled by how long they remain in the high density shock front. While core mergers can play a role in increasing core mass, the majority of the mass is gained via accretion. This happens most efficiently while the background density is high; that is, while the core is inside the shocked disc. If the shock moves too fast, the cores exit the disc before they have had enough time to gain sufficient mass to become bound. These two processes (core formation and core growth) dictate the ideal shock speed for core production: fast enough to allow significant core formation but slow enough to give these cores time to accrete. 

The production of potentially star forming gas is seen in the cloud PDFs. With an initially turbulent structure, the clouds begin with a log-normal profile at high densities. However, when core formation starts, a non-log-normal tail develops. This agrees with observational results by \citet{Kainulainen2009} who observed that GMCs possessed such a tail once they began forming stars. A similar result was also found in Orion B by \citet{Schneider2013}. This evolution in the PDF is intricately linked with the production of turbulence in the clouds. As the shock front forms between the clouds, compressive turbulence modes are formed which increase the star formation efficiency and lead to the PDF tail.

The cumulative mass distribution for the cores fit a power law with $\gamma = -1.6$ (from Equation~\ref{eq:cmd}), in good agreement with observed cores inside GMCs. This suggests that cores formed in cloud collisions may resemble those in quiescent star formation, although the increased density of the shocked disc would be expected to increase both initial core size and accretion speed. Indeed, if allowed to collapse without colliding, our clouds only form a small number of low density ($\rho = 5\times 10^{-21}$\,gcm$^{-3}$) cores and do not produce objects close to the observed density of star-forming regions \citep{Lada2010}. 

The fact we produce substantially more massive cores in the collision than are formed through gravitational collapse alone is an indication that this process may be important for massive star formation. However, since accretion is the dominant forms of core growth in our colliding clouds, the resulting stars are unlikely to exceed 10\,\msun. Two possible effects may change this result. The first is the size of our clouds, which are at the lower end of the observed GMC mass distribution. A collision between larger clouds or ones with initially higher densities may result in more massive fragmentation, producing larger cores prior to accretion. The second possibility is that cores are prevented from collapse by magnetic fields or turbulence that we are not resolving at our current resolution. This would permit the cores to grow via accretion before radiation from the forming star cuts the gas flow. We will consider these effects in later work.

\section*{Acknowledgments}

\acknowledgments

The authors would like to thank Yasuo Fukui for many helpful discussions about this work, the yt development team \citep{yt} for support during the analysis of these simulations and their anonymous referee for an extremely helpful report. Numerical computations were carried out on the Cray XT4 and Cray XC30 at the Center for Computational Astrophysics (CfCA) of the National Astronomical Observatory of Japan. EJT is funded by the MEXT grant for the Tenure Track System.

\end{document}